\documentclass[preprints,article,accept,moreauthors,pdftex]{Definitions/mdpi} 

\newcommand{\mt}{\emph{MysTail}}
\firstpage{1} 
\pubvolume{1}
\issuenum{1}
\articlenumber{0}
\pubyear{2021}
\copyrightyear{2020}
\datereceived{} 
\dateaccepted{} 
\datepublished{} 
\hreflink{https://doi.org/} 

\Title{One Source, Two Source(s): Ribs and Tethers}

\TitleCitation{One Source, Two Source(s)}

\Author{Lawrence Rudnick $^{1,\dagger}$, William Cotton $^{2}$, Kenda Knowles $^{3,4}$ and Konstantinos Kolokythas  $^{5}$}

\AuthorNames{Lawrence Rudnick, William Cotton, Kenda Knowles and Konstantinos Kolokythas}

\AuthorCitation{Rudnick, L.; Cotton, W.; Knowles, K.; Kolokythas, K.}

\address{
$^{1}$ \quad Minnesota Institute for Astrophysics, University of Minnesota, 116 Church St SE, Minneapolis, MN 55455, USA; larry@umn.edu\\
$^{2}$ \quad National Radio Astronomy Observatory, Charlottesville, VA 22903, USA\\
$^{3}$ \quad Department of Physics and Electronics, Rhodes University, PO Box 94, Makhanda 6140, South Africa\\
$^{4}$ \quad South African Radio Astronomy Observatory, 2 Fir Street, Observatory, Cape Town 7405, South Africa\\
$^{5}$ \quad Centre for Space Research, North-West University, Potchefstroom, 2520, South Africa\\
}
\corres{Correspondence: larry@umn.edu; Tel: 1-612-624-3396}

\firstnote{Current address: Minnesota Institute for Astrophysics, University of Minnesota, 116 Church St., SE, Minneapolis, MN 55455, USA} 

\abstract{We present the unique and challenging case of a radio galaxy in Abell~3266 observed as part of the MeerKAT Galaxy Cluster Legacy Survey.  It has quasi-periodic bright patches along the tail which connect to never-before-seen thin transverse extensions, which we call `ribs', reaching up to $\sim$\,50~kpc from the central axis of the tail. At a distance of $\sim$\,400~kpc from the host (assuming the $z=0.0594$ redshift of Abell~3266) we find what appears to be a triple source with its own apparent host at a photometric redshift of 0.78.  Mysteriously, the part of the tail far from the host and the triple are connected by a series of thin filaments, which we call ``tethers.'' The far tail, tethers and triple also have similar spectra and Faraday rotation measures, suggesting that there is only one -- quite complicated -- source, with a serendipitous background AGN in the triple.  We look at possible causes for the ``rib'' and ``tether'' structures, and the emerging phenomena of intracluster medium filaments associated with radio galaxies. }


\begin{document}

\section{Introduction}
The study of radio galaxies is entering a new phase, with the exciting capabilities of SKA precursors such as MeerKAT and ASKAP, low frequency imaging at high resolution with LOFAR, and the high sensitivity, resolution and broad frequency coverage of the Jansky VLA.  We are finding not only new examples of astrophysics -- e.g., how radio galaxies interact with their environment  -- but new phenomena that may require revision of our basic theoretical pictures of jet physics.  Here, we present one such case, which invites us to revisit what we assume we know about jets and their tails.

\begin{figure}[h]
\centering
\includegraphics[width=0.9\columnwidth]{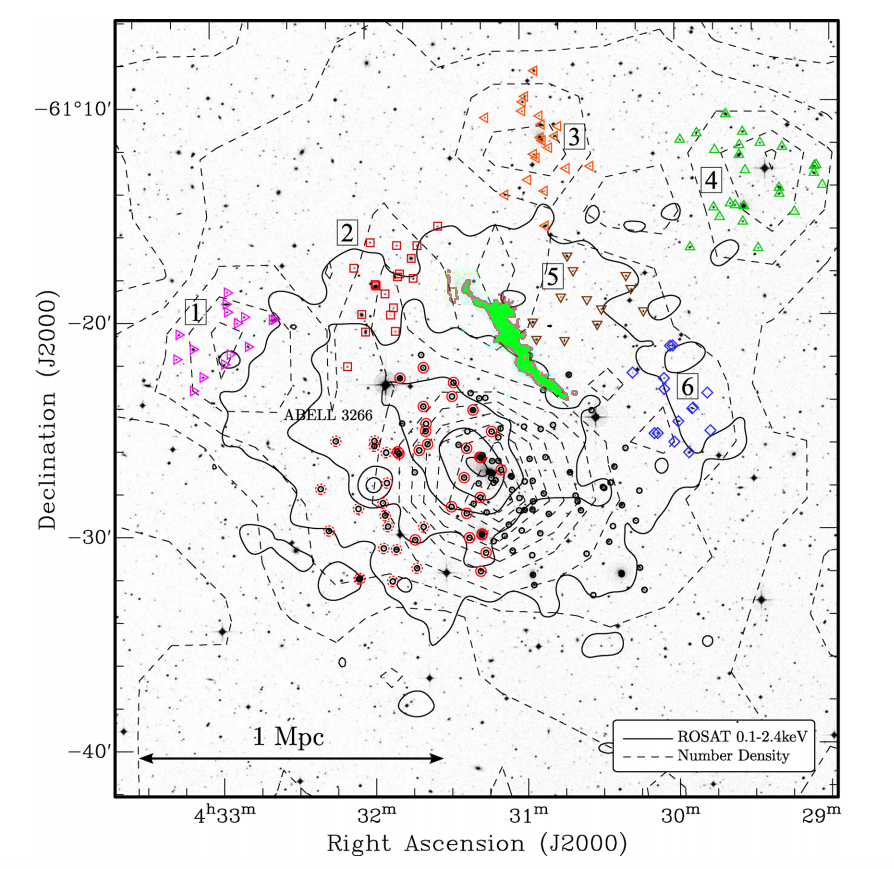}
\caption{Distribution of optical galaxies in Abell~3266 from \citet{D17}, (their Figure 7), showing the location and size of \mt\ in light green.}
\label{cluster}
\end{figure}

\begin{figure}
\centering
\includegraphics[width=0.9\columnwidth]{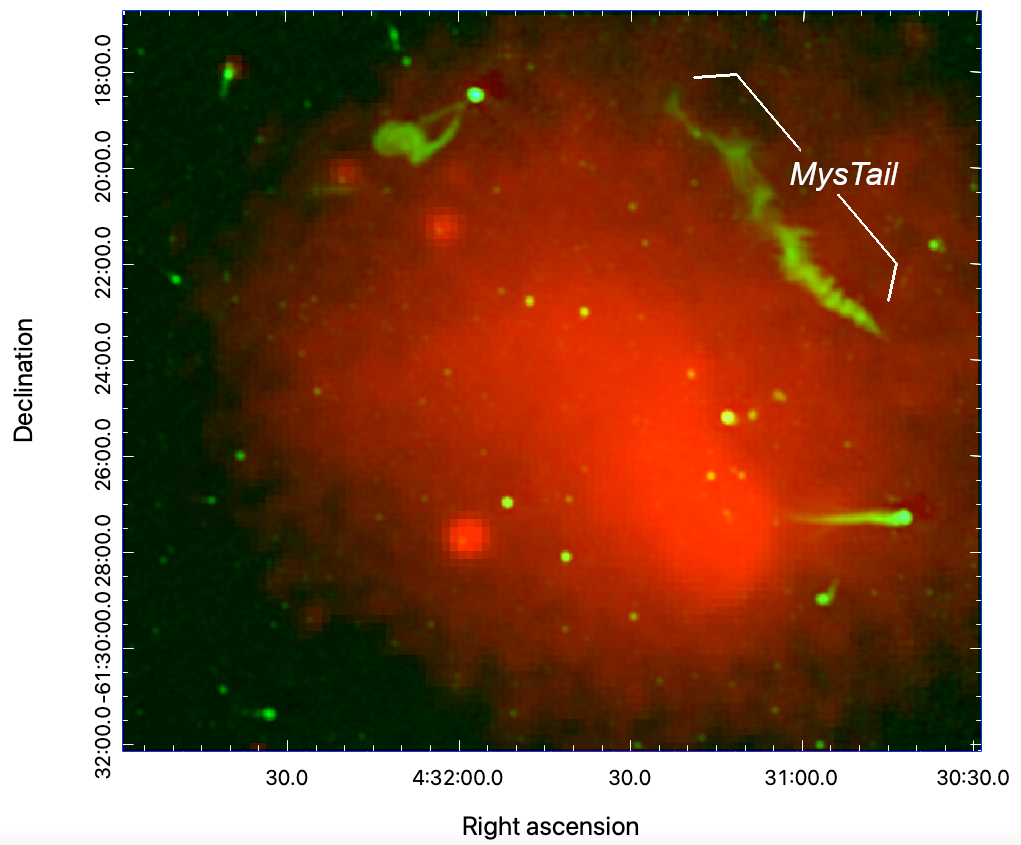}
\caption{Radio map in green of the central $\sim$\,15 arcmin of the cluster region, overlaid with a \textit{XMM-Newton} 0.7--2.3~keV image courtesy of Hiroki Akamatsu in red.  
}
\label{radioX}
\end{figure}

The mysterious radio galaxy in question, which we hereinafter call \mt, is a member of Abell 3266.  Initial results on this source have been presented in \citet{MGCLS}. The cluster is part of the Horologium-Reticulumin supercluster, and is in the midst of a complex merger, with velocities spanning 14,000 to 22,000~km/s over a radius of 1$^{\circ}$, or $\sim$\,4~Mpc at the central cluster redshift of 0.0594 \citet{D17}.  The core region of the cluster, shown in Figures \ref{cluster} and \ref{radioX}, has two major  components separated along a  NE to SW axis, with a number of other filaments and subclusters to the north.  \citet{D17} suggest that the ongoing merger activity suggests a high level of turbulence in the intracluster medium (ICM).   The location of \mt\ relative to the cluster is shown in Figure \ref{cluster}. \mt\ is elongated at the edge, and along the same axis, as the major velocity components in the cluster core.  This structure was previously seen by \citet{bern16} with the MeerKAT precursor, KAT-7, although they were unable to classify its nature.

The X-ray structure (\citet{fin06}) is elongated along the same NE--SW axis as the optical core, with unusual low entropy gas that they suggest has been stripped from an infalling subcluster.  Recent eROSITA observations (\citet{Sanders21}), show an X-ray structure and pressure discontinuity further to the northwest, and other signs of the complex merger and possible AGN perturbations.  An overlay of the radio image with the \textit{XMM-Newton} X-ray image of the inner $\sim$\,15$^\prime$ of the field is shown in Figure \ref{radioX}.  Other interesting radio galaxies and radio relic structures are also present in the field, but are beyond this image and the scope of this work.

\section{OBSERVATIONS and OVERVIEW}
The observations presented here were carried out as part of the MeerKAT Galaxy Cluster Legacy Survey (MGCLS; \citet{MGCLS}).  Details of the observations and map production are given there, but we include a brief summary here. 

The observations were made with the MeerKAT array on 2019 March 3 for 8 hours using J0421$-$6223 as the gain calibrator and PMN~J0408$-$6545 as the flux density/bandpass calibrator.
The observed integration time was 8~seconds and with 0.21~MHz channels covering the nominal frequency range 900 -- 1670 MHz.
The primary antenna beam at mid--band is 67$^\prime$ FWHM and the resolution of the image is 7.1$^{\prime\prime}$ $\times$ 6.7$^{\prime\prime}$ at position angle 31$^\circ$.
All four combinations of the two linear feeds were correlated.

Polarization calibration was as described in \citet{MGCLS} and used
the noise signals injected at the beginning of each observing session
and calibration parameters derived from external data sets.
Imaging in Stokes Q and U used the same 5\% fractional bandwidths as
were used for Stokes I. 

After calibration the data were averaged to 0.83 MHz channels for imaging.
At the location of \mt, 350$^{\prime\prime}$ from the field center, the RMS is 6~$\mu$Jy beam$^{-1}$.
The spectral index $\alpha$, (using the convention $S(\nu)=S_0~\nu^{\alpha}$), was fitted in each pixel using the sub--band images derived in the imaging after primary beam correction.
\subsection{Dissecting the structure} 
\begin{figure}
\centering
\includegraphics[width=0.9\columnwidth]{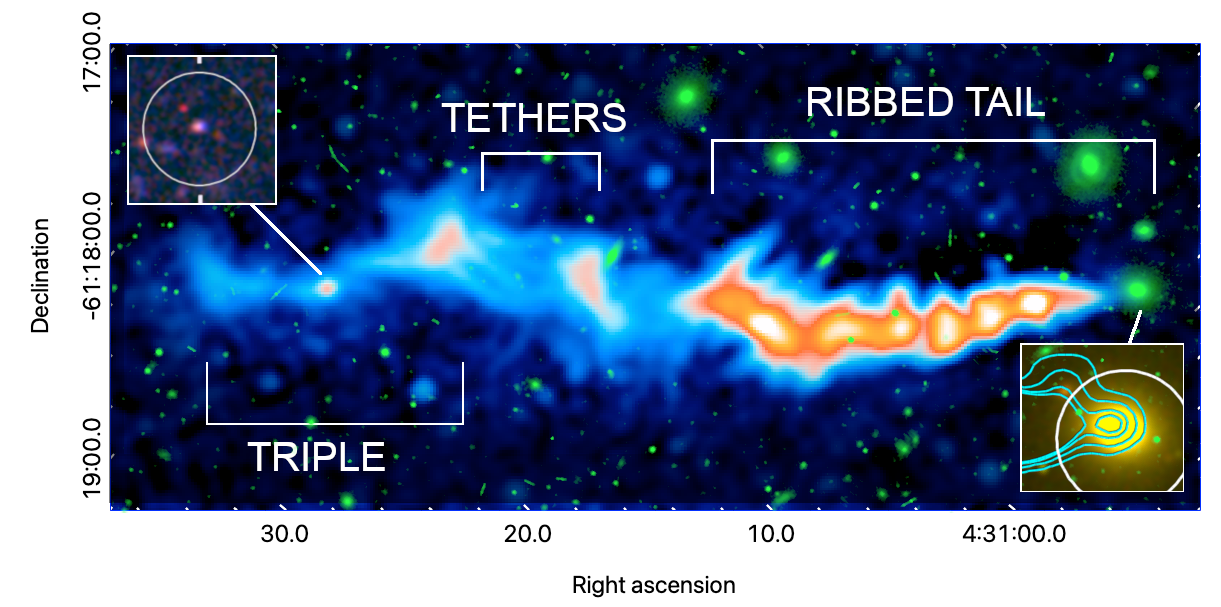}
\caption{MGCLS 1285~MHz radio map of \mt\ with suggested host(s) and labeled features. Background green image and insets show the optical field from the Dark Energy Survey. Radio contours on the inset indicate the presence of a core coincident with the suggested host.}
\label{radioID}
\end{figure}

Figure \ref{radioID} shows a close-up view of \mt, indicating the structures discussed here and the proposed optical identifications.  At present, we cannot determine whether we are looking at a single radio structure, or multiple structures either physically linked or seen only in projection.  Another alternative is that \mt\ could be a peripheral radio relic, because it follows the elongated X-rays in the cluster core. eROSITA studies show a parallel pressure jump structure, although it is much further out.  The identification of \mt\ as a tailed radio galaxy, however, is indicated by the existence of a radio core  and coincident bright galaxy at one end (see right-most inset in Figure \ref{radioID}), coupled with the spectral behavior described below. The optical galaxy is WISEA~J043045.39$-$612335.6 (6dFGS g0430454-612336), at a redshift of 0.0626, within the velocity distribution of the Abell~3266 core.   The tail is resolved transversely ($\sim$\,20$^{\prime\prime}$ wide in the faint regions, with a 7.2$^{\prime\prime}$ beam), but shows no signs of bifurcating into a twin-tailed structure.  It is punctuated by quasi-periodic bright patches, and shows much more extended transverse structures, reaching over an arcminute in extent, which we label ``ribs''.  At 2.1$^{\prime}$ ($\sim$150~kpc) from the core, the tail shows a sharp bend of $\sim$35$^{\circ}$. It continues for $\sim$\,40$^{\prime\prime}$ before bending again and getting dimmer.  A bright patch appears $\sim$50$^{\prime\prime}$ further downstream.

At the far (NE) region of the source is a structure which we label as the ``triple'', because of its similarity to a double-lobed radio galaxy with a  core and connecting jets. Approximately half of the 0.64~mJy\,beam$^{-1}$ core brightness may come from the underlying jet.  In addition to the cluster radio galaxies, there are approximately 15 compact radio sources with peak fluxes $>$\,0.3~mJy\,beam$^{-1}$, within 1.2$\times10^5$ square arcseconds; the chances of being located somewhere within the jet-like structures is thus less than a few percent. Coincident with the presumed core, we find a faint optical/IR galaxy (DES~J043118.45$-$611917.9 WISEA~J043118.50$-$611918.2) with a photometric redshift of 0.78 (\citet{2019ApJS..242....8Z}). At this redshift, the 120-arcsec possible triple source would be 900~kpc in length; if the triple were an extension of the tail, it would be 140~kpc long.  Between the bright patch at the end of the tail and the triple we find three filamentary features, which we label as ``tethers''.  The connection between all of these structures is unclear; hence the interest in this source.  

\begin{figure}
\centering
\includegraphics[width=0.9\columnwidth]{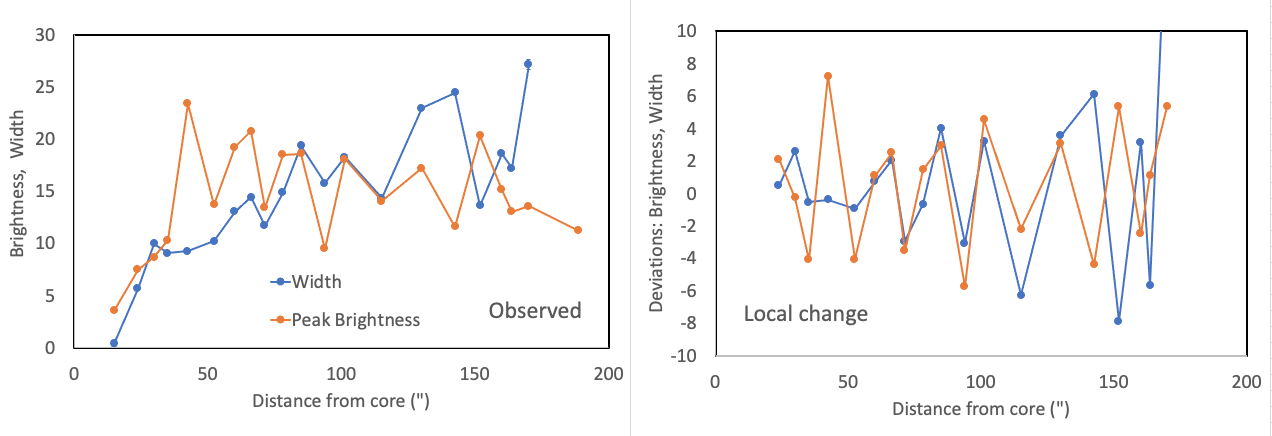}
\caption{Gaussian fits of peak and deconvolved width to profiles across the tail. \textbf{(Left)} as observed, in units of 10$^{-4}$ Jy (peak), and arcsec (width); \textbf{(Right) } difference between each point and the average of its two neighbors, in units of  (\% deviation)/10 (peak) and arcsec (width). Statistical error bars are approximately the size of the symbols; the true uncertainties are dominated by the non-Gaussianity of the profiles and the background subtractions.}
\label{peakwidth}
\end{figure}

\section{Source diagnostics}

\subsection{Structure}
In order to look at the properties of the bright spots and ribs in more detail, we made transverse cuts at a number of locations along the tail and measured their peak brightnesses and Gaussian widths.   Note that these width measurements are only approximate indicators, and include the broadening by the ribs. A more accurate width determination would require assumptions about the true underlying brightness profiles of the tail and ribs.  The quasi-periodic brightening of the tail is seen in Figure \ref{peakwidth}. The characteristic separation between the bright spots is $\sim$\,20$^{\prime\prime}$, comparable to the width of the tail.  The tail widens over the first 200$^{\prime\prime}$, past its initial bend. The ribs are shorter near the head, and appear mostly on one side, reaching $\sim$\,20~kpc from the center line of the tail.  The longest rib is seen at the end of the bend, and spans a total length of $\sim$\,100~kpc.  The widths of the ribs range from about 4 -- 7 kpc, after beam deconvolution.

In order to look more closely at the small-scale variations, we took out the large-scale trends by taking the difference between each peak and width and the average of the two neighboring measurements. The results are also shown in Figure \ref{peakwidth}.  The key result here is the same as can be seen in the image;  the ribs often occur at the bright patches in the tail, although this relationship gets confused around the bend, where the structures become more difficult to match up with one another.

\subsection{Spectral index variations}

\begin{figure}
\begin{center}
\includegraphics[width=0.9\columnwidth]{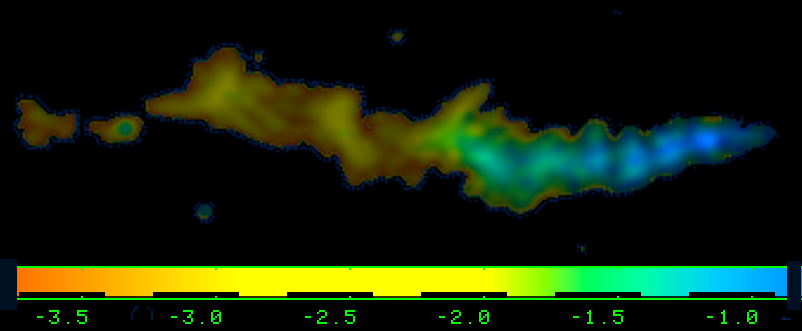}
\end{center}
\caption{Spectral index color coding of total intensity map.}
\label{huein}
\end{figure}

The overall spectral index behavior, shown in Figure \ref{huein}, is similar to what is seen in other tailed radio galaxies.  The index is $-0.6$ near the core, steepening to $\sim$\,$-1.4$ at the bend in the tail. Beyond the second bend, there is a dramatic steepening with values ranging from $\sim$\,$-2$ to $\sim$\,$-3.4$  (except for the putative core of the triple, with $\alpha=-1.4$).  We did several spot checks of the pixel-by-pixel spectral fits by integrating over larger areas in each of the frequency channel maps and refitting for the spectral index.  The results were consistent. Note that the ribs follow the same overall spectral trends as the tail.  

In addition, we find no evidence for spectral curvature, even when the spectra were steep, with changes in spectral index less than 0.5 between the top and bottom halves of the band.  For comparison, an exponentially cut off spectrum with a lower frequency index of $-0.6$ would have steepened by $\sim$1.5.  

 \begin{figure}
\centering
\includegraphics[width=0.8\columnwidth]{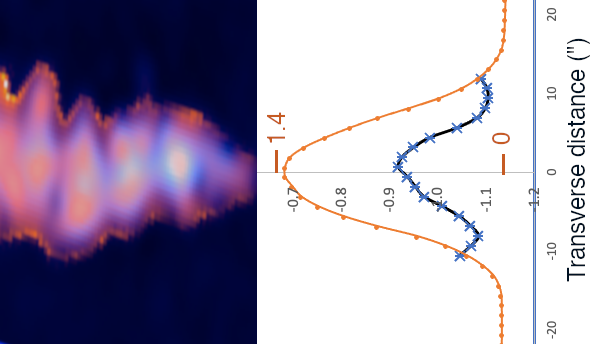}
\caption{Closeup of beginning of tail to highlight spectral index structure, stretched vertically by a factor of 2. {\em Left}: Spectral tomography images.   Blue (orange) shows emission flatter than (steeper than) $\alpha$= -0.63. These cannot be used quantitatively; they simply illustrate the overall spectral structures. {\em Right}: Profile cuts across the tail in total intensity (orange) and spectral index (blue); these were produced by taking the median value of the respective images along the $\sim$1' portion of the tail shown here, at each distance from the tail central line.}
\label{ialphacut}
\end{figure}
A closer look at the spectral structure is shown in Figure \ref{ialphacut}.  This image was made using ``spectral tomography'' which helps identify overlapping structures with different spectral indices (\citet{tomo}).  There appear to be two parallel lines of slightly steeper spectrum material surrounding the ridge line of the tail, although the flatter spectrum emission extends transversely beyond those lines.  This behavior can be seen in the profile plot, which shows a steepening away from the ridge line, with some modest flattening further out.  Note that the total intensity cut blends all of these features together.  The true underlying structure and spectral decomposition would require resolutions $<$\,3$^{\prime\prime}$;  the results here show simply that the tail cross-section is not uniform.

\subsection{Tethers} 
\begin{figure}[H]
\begin{center}
\includegraphics[width=12 cm]{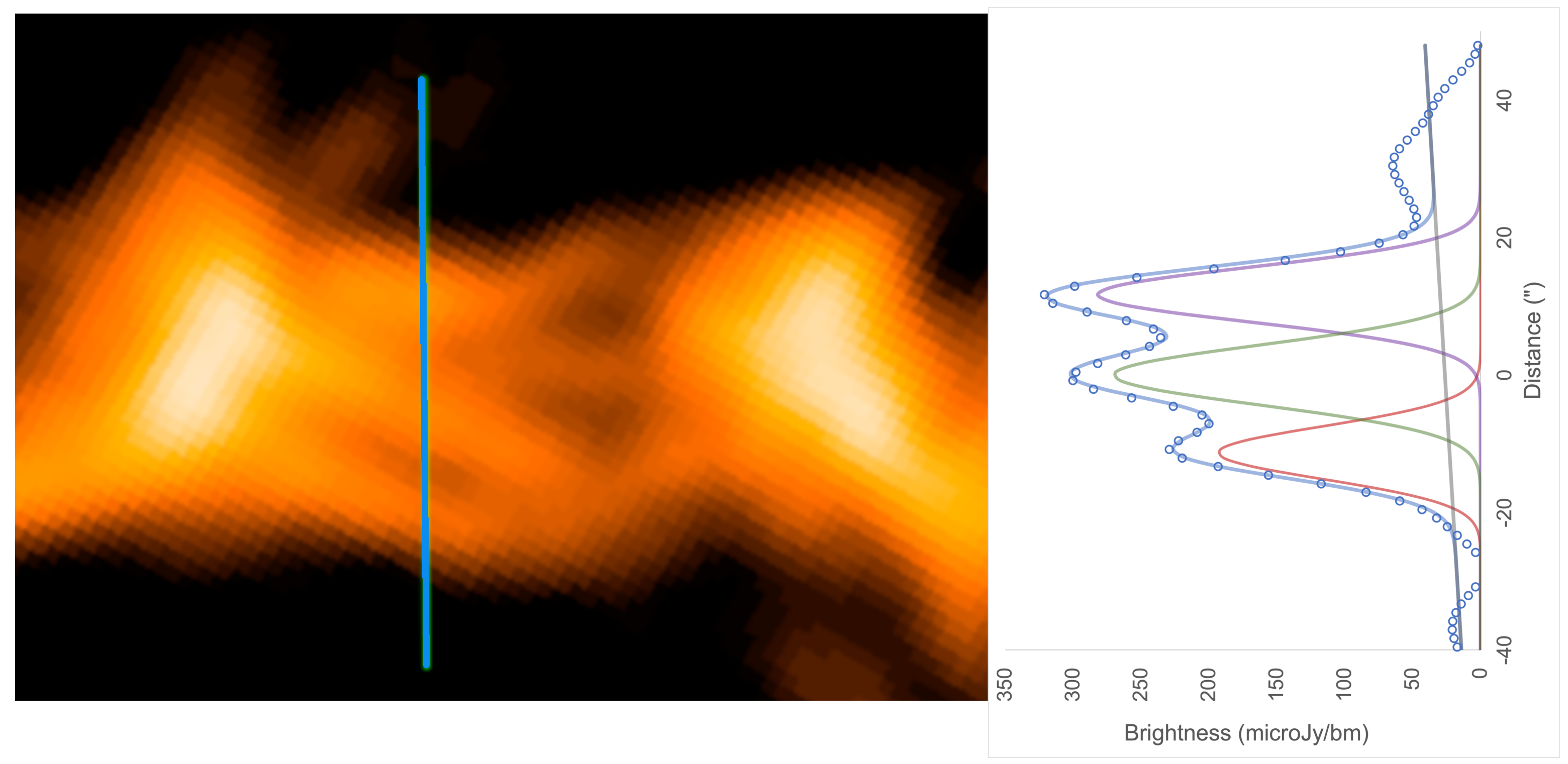}
\end{center}
\caption{Closeup of the tethers region, with a profile cut showing that it can be characterized as three somewhat resolved Gaussians.  }
\label{teth}
\end{figure}

The tethers have an average brightness $\sim$\,250$~\mu$Jy/beam, comparable to the ribs, the fainter portions of the tail, and the triple.  Their length is $\sim$\,45$^{\prime\prime}$  (52~kpc, assuming they are associated with Abell~3266), and their widths are slightly resolved, with deconvolved values of 6--7$^{\prime\prime}$ (7--8~kpc).  Figure \ref{teth} shows a profile cut across the tethers, with a 3-Gaussian model.  The blending of three slightly resolved Gaussians accounts for all the flux in the tether region, although it does not rule out a combination of multiple thinner structures and more diffuse emission that could become visible at higher resolution.  A faint partial fourth tether is also visible at some locations. Cross sectional cuts are also consistent there with three Gaussians, with the fourth tether replacing one of the two merged brighter tethers.

\subsection{A first look at polarization}
The Faraday rotation in the field was derived from a Faraday synthesis
of the Q and U image cubes.  A direct search was made in each pixel
between +300~rad~m$^{-2}$ and $-300$~rad~m$^{-2}$ with increments of 1 rad~m$^{-2}$.
The value of rotation measure that produced the maximum unwrapped
polarized emission was taken as the rotation measure of that pixel. The polarized intensity (at the peak in the Faraday spectrum) and the electric vector polarization angle corrected for Faraday rotation were then calculated.

\begin{figure}[H]
\begin{center}
\includegraphics[width=0.9\columnwidth]{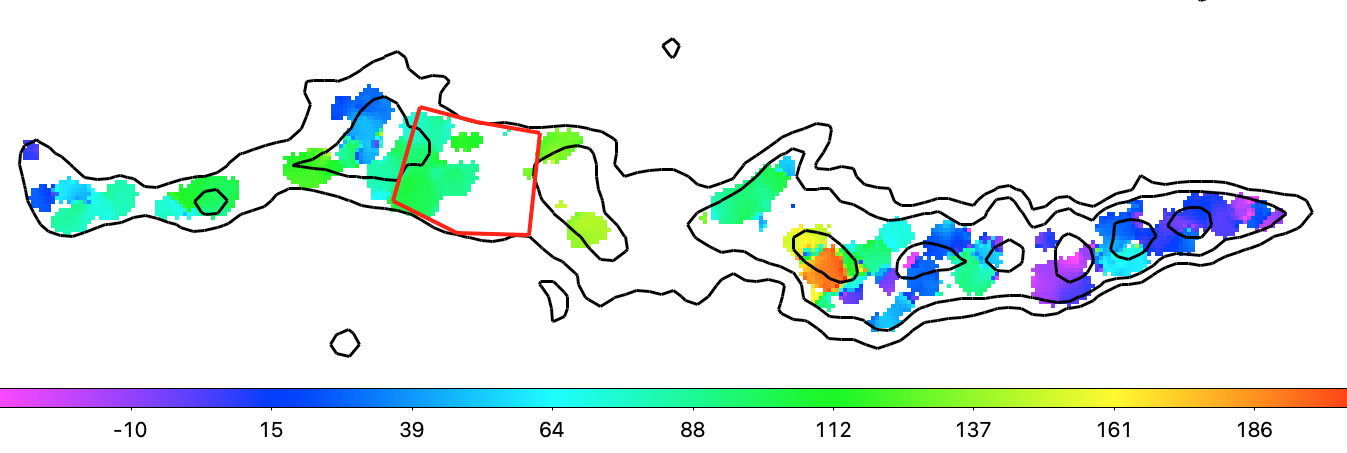}
\includegraphics[width=0.9\columnwidth]{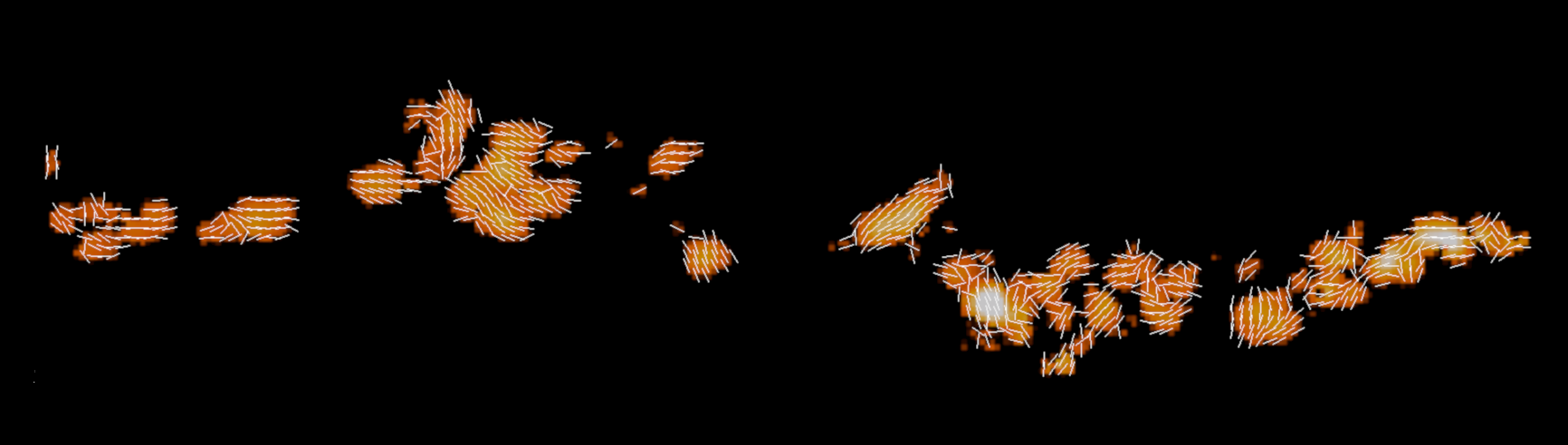}
\end{center}
\caption{{\bf Top:} Rotation measure, with scale in rad~m$^{-2}$ with total intensity contours spaced by factor of 4, starting at 81 $\mu$Jy\,beam$^{-1}$.  The red polygon denotes the region of the tethers.{\bf Bottom:} Heatmap of polarized intensity ranging from 25 - 150 $\mu$Jy/beam, superposed by inferred magnetic field direction vectors after correction for Faraday rotation. }
\label{RM}
\end{figure}

The results are shown in Figure \ref{RM} where the signal:noise was high enough to determine a reliable RM.
In the initial 125$^{\prime\prime}$ section of the tail, prior to the bend, the average RM is 22 rad~m$^{-2}$, with an RMS scatter of $\sim$30 rad~m$^{-2}$. The polarization is seen to be patchy, as is the RM, with no obvious systematic orientation to the inferred magnetic field vectors. The fractional polarization, where it can be measured, averages $\sim$7\%.

There is a sharp transition in the RM values at the bend, with a mean $\sim$\,120~rad~m$^{-2}$, and scatter of $\sim$\,45~rad~m$^{-2}$. In general, the RM variations are on larger scales and the mean fractional polarization rises to $\sim$11\%.  In addition, there is a compact region of high polarized flux at the bend, and a large jump in RM. These suggest some interaction with the surrounding thermal material. There is no significant change in the spectral index of $\sim$-1.4 at this position.

The RMs remain at similar high values in the region of the tethers, (95\,$\pm$\,10~rad~m$^{-2}$) and the triple (75\,$\pm$\,35~rad~m$^{-2}$), again with variations over larger scales than the initial section of the tail. The mean fractional polarization is higher still, $\sim$25\%. The magnetic fields in the large extended features, including the tethers, tend to align with the structures.  

Significant changes in the polarization characteristics (RMs, field directions and fractional polarizations) are thus seen to correlate with the overall source structure.  Interactions with the surrounding medium are likely to play an important role in this behavior.

\section{Discussion - what are we seeing?}
\subsection{Bright patches and ribs}
We briefly summarize the key features of these structures as follows:
\begin{itemize}
    \item{A radio tail with no visible bifurcation, with quasi-periodic brightness variations over its initial 200~kpc.}
    \item {The brightness modulations are a factor of $\sim$\,1.5 with a separation between  peaks  on the order of the 2 times the tail diameter. }
    \item{At the positions of the bright peaks, thin ($\sim$\,5 ~kpc)  transverse ``ribs'' extend up to $\sim$\,100~kpc in length across the tail.}
    \item{All of the structures have spectra which steepen with distance down the tail, with the ribs typically locally steeper by about 0.2 than the core of the tail.}
\end{itemize}

We note that quasi-periodic jet brightenings have been seen previously, including the spectacular example of PKS~0637-752  (\citet{godfrey}) with $>$\,10 knots spaced $\sim$\,7.6~kpc apart (in projection). They interpret these as due to variations in jet output on timescales of 10$^3$--10$^5$ years. Although no transverse structures were noted, deeper observations of this system are certainly warranted.  

In general terms, there are three plausible ways in which these structures could arise:   variations in the AGN jet output; instabilities in the flow (including possible interactions with an inhomogeneous external medium );  coherent pre-existing external structures superposed on/ interacting with the tail.  We provide brief initial speculative thoughts on these possibilities.

\ul {Pulsed/varying jets?} Temporal variability of flux and structure in jets is well-studied  on pc-scales (\citet{Lister}). At the other extreme, episodic jet activity has been proposed to explain much larger scale structures in double-double radio galaxies, and the curious multiple structures in a spiral radio galaxy (\citet{Sirothia} ).  On kpc and larger  scales,  simulations with various jet duty-cycles have been used to study the feedback effects on the surrounding cluster medium or on galaxy formation (\citet{feedback}).  Re-starting jet studies reveal rich underlying physics, such as the differences between jet injection into pre-existing channels from previous activity or jets that need to excavate a new path (\citet{restart}) .  For \mt, jet variability could be at play if the outbursts were highly overpressured and lead to transverse ``ribs'';  however, this has yet to be seen or predicted in simulations, so further work is needed. In addition, the spacing of the bright patches on similar scales to the jet diameter, would be fortuitous.

\ul {Instabilities in the flow -- recollimation shocks?}
Various instabilities naturally develop in flows, largely through interactions with the external medium (\citet{oneill}).  These can result in flapping-type motions, entrainment of external material, and the creation of vortex rings when shocks are encountered (\citet{noltingv}).  One interesting possibility is whether recollimation shocks are present; they can be created when jets over-expand and are then reconfined as they propagate into lower density environments (\citet{Sanders}). \citet{Mizuno} has also explored the role of axial magnetic fields in pc-scale jets in strengthening recollimation shocks and the resulting compressions and rarefactions. Recollimation processes such as these might therefore be an attractive possibility for creating the bright knots, but the broadening of the jets at the bright spots, and the development of ``ribs'' does not appear consistent with this process.

\ul {External (ICM) structures? }
Pre-existing density fluctuations, or especially, filamentary structures, in the ICM are attractive candidates for the ribs, and are discussed further below in the context of the ``tethers''.  We also note the large, thin transverse features found in Centaurus A, and associated with bright knots in its Northern Middle Lobe (\citet{mck}).  In that case, the interaction of strong mildly supersonic winds from the AGN appear to be interacting with clumps of X-ray emitting material, giving rise to these features.  There is no indication or expectation that this particular mechanism would be applicable to \mt.

\subsection{Tethers --  connecting one or two sources?}
To interpret the tethers, we must first determine whether the tail, which itself is clearly part of Abell~3266,  is physically associated with the triple.  Either way, we are faced with an unsettling coincidence.  If these are one physical source, then the optical/radio ``core'' must be an accidental background source. We calculated a $\sim$\,3\% chance that such an object would be found somewhere along the apparent jet-like structures in the triple.  If we also note the approximate central position of the core, the probabilities would go down by another factor of 5. 

\begin{figure}
\begin{center}
\includegraphics[width=0.9\columnwidth]{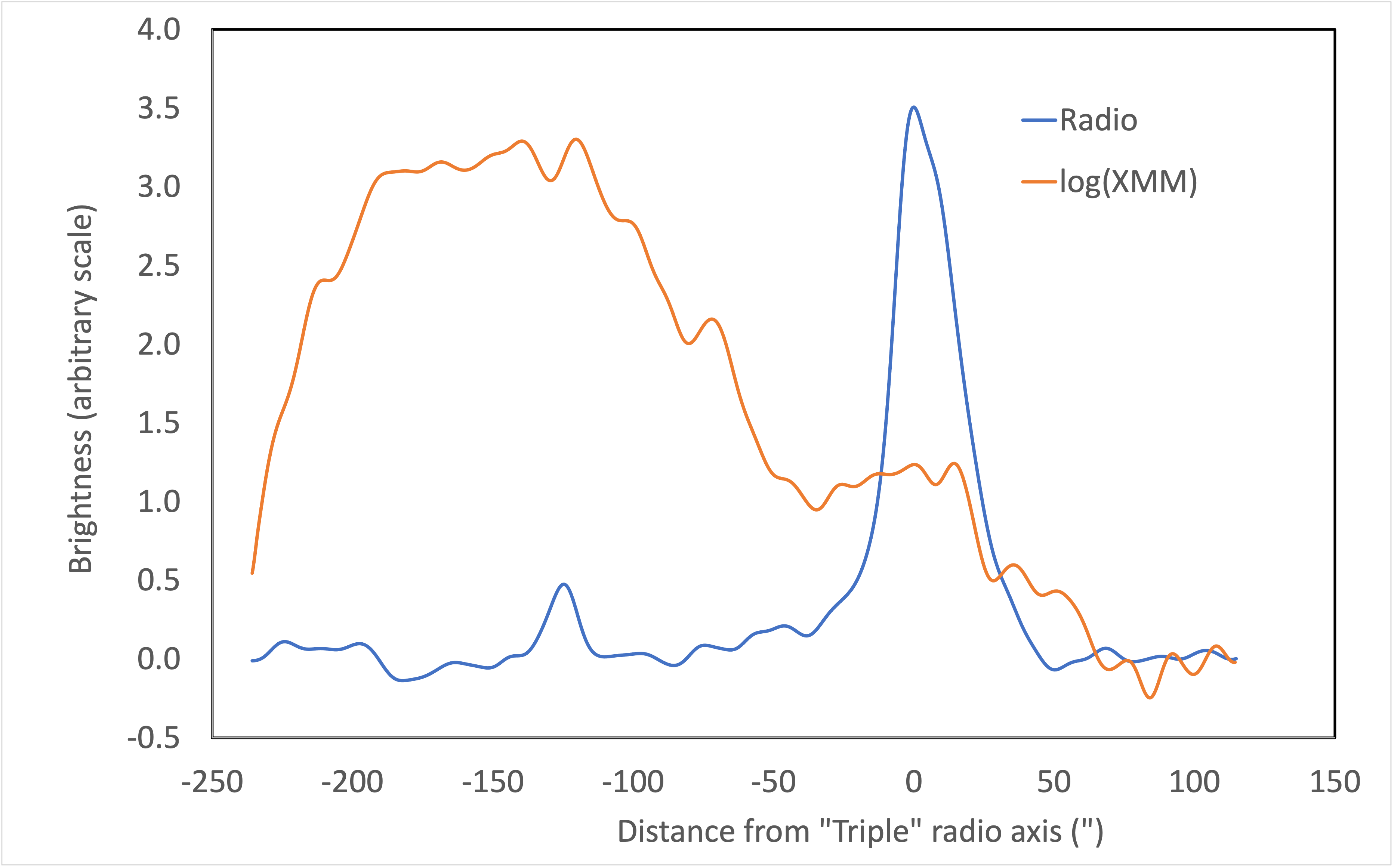}
\end{center}
\caption{Brightness slices in radio and X-ray (from XMM) perpendicular to the ``triple''. These were produced by first convolving the images with a Gaussian of 125$^{\prime\prime}$ $\times$ 10$^{\prime\prime}$ at an angle of 44deg, along the triple structure.  The brightness scales are arbitrary, and in log brightness for the X-rays. These plots show that the triple is located at/near a drop in X-ray brightness.}
\label{xt}
\end{figure}

At present, accepting this coincidence appears preferable to the coincidences we'd have to accept if these are two separate sources, seen in projection.  These include:
\begin{itemize}
    \item {That the tail and triple fall along the same line, and appear connected by thin filaments (the tethers).}
    \item{That the spectral index smoothly steepens from the tail, through the tethers, and from the west to the east through the triple. The very steep spectra in the triple would also be unusual for a random background source, although more typical for the faint regions of a tail.}
    \item{That the RM distribution in the triple is similar to that of the tethers and the post-bend regions of the tail, and finally;}
    \item{That the triple is parallel to, and  appears to be located at/near a drop in the cluster X-ray brightness, as shown in Figure \ref{xt}; note that this drop is located interior to the pressure jump seen with eROSITA (\citet{Sanders21}).}
 \end{itemize}
 
\begin{figure}[H]
\begin{center}
\includegraphics[width=11 cm]{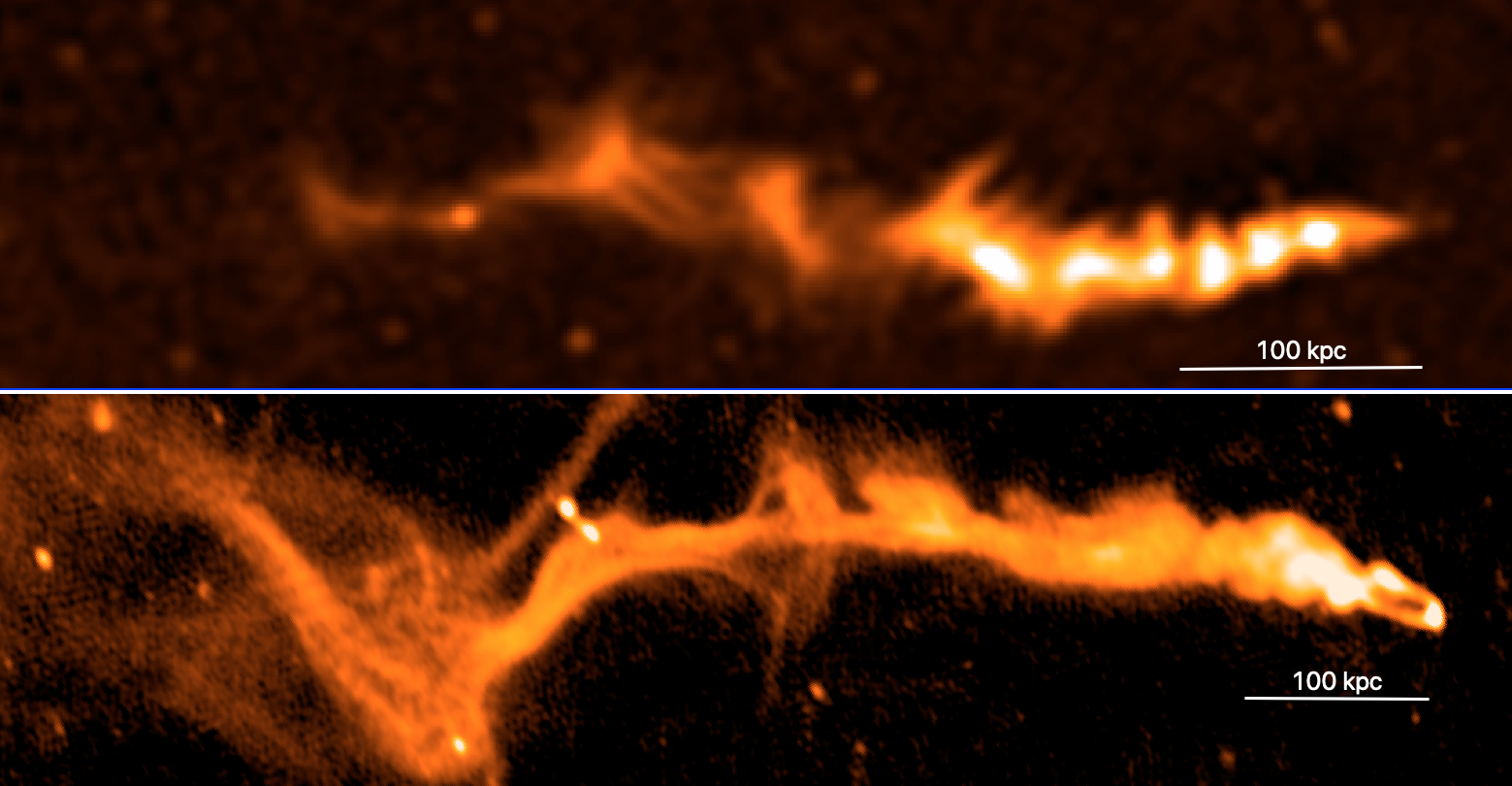}
\end{center}
\caption{MeerKAT observations of \mt, as presented here, and image produced using LOFAR observations of IC711, courtesy of \citet{vanW21}, on similar scales, showing the emergence of heretofore unusual features associated with tailed radio galaxies. }
\label{compare}
\end{figure}

We therefore adopt, for the purposes of discussion, that we are looking at one source comprising the tail, the triple, and the connecting tethers.  Other than the unusual nature of the ribs, as discussed above, the tail is similar to many other tailed radio galaxies that have been studied over the years.  The triple structure, the somewhat isolated bright patch between the tethers and the tail and the tethers themselves are quite unusual -- but not unique.   Similar features, e.g., are seen in  IC711 (\citet{vanW21}), with patchy structures, multiple filamentary structures intersecting with the tail, possibly connecting to other sources in the host cluster Abell~1314,  etc. (see Figure \ref{compare}). The interactions of radio tails with the ICM are likely much richer than we have observed or modeled through simulations, up until now.

Filamentary features associated with radio galaxies are becoming more common.  An extensive network of filaments was found in Abell~194, also observed as part of the MGCLS Survey (\citet{MGCLS}), around the radio galaxies 3C40A,B.  \citet{threads} show multiple ``collimated synchrotron threads'' connecting the lobes of radio galaxy ESO~137$-$006. \citet{condon21} describe the ``threads'' and ``ribbons'' in IC4296, which they suggest are related to instabilities in the jets and relics of earlier jet activity, respectively. Even the well-studied radio galaxies 3C129, NGC~326 and IC711  have revealed strange filamentary features connecting with their more conventional structures \citet{lane,hardcastle,vanW21}.

Filamentation may thus occur naturally in the ICM, with the magnetized filaments becoming populated, and thus illuminated by relativistic particles when reconnection takes place with radio galaxy lobes.   Similar processes may play out in our own Galaxy, where massive winds in the galactic center interact with strong fields in the interstellar medium, leading to equally spaced structures called ``synchrotron harps'' by \citet{thomas}. Such structures could also be responsible for the ``ribs'' discussed above. If such filaments intersected with the tail, and magnetic field reconnection took place, it is possible that they would become illuminated by relativistic electrons from the tail. Diffusion times, spectral expectations, lengths and widths -- these are all aspects that would need study to make such a scenario plausible.

\subsection{The tail -- X-ray connection}
Radio galaxies are clearly distorted by their interactions with the ICM and, in the dramatic case of encountering shocks, may be responsible for forming peripheral radio relics (e.g., \citet{jones21}).  Although there is no shock structure apparent at the location of \mt, it does lie along the edge of the low-entropy bright central cluster emission, and its structure could be associated with flows along or pressure gradients across this edge. The spectral gradients along its length suggest that the relativistic electrons are still evolving passively as expected down a tail, but the unusual structures might indicate interactions with the ICM. The sharp bend in the tail at $\sim$150~kpc from the core is accompanied by changes in the Faraday and magnetic field structure;  these are likely signatures of a change in the physical and/or dynamical properties of the local ICM.

\section{CONCLUSIONS}
In the end, \mt\ presents several mysteries, with properties that do not fit cleanly into our current models of radio galaxies. It is not clear whether this is even a standard twin-tail source, with the tails merged in projection or physically, or whether it is new type of one-sided structure.  While a variety of factors could be responsible for the bright patches down the tail and the accompanying ``ribs'', none of them satisfactorily accounts for all the observed properties.  Added to this puzzle, the continuity of structure, spectrum and rotation measure from the far tail, through the tethers, to the triple source, suggest that they are one structure, but the apparent host at a different redshift must then be viewed as an improbable coincidence.  If we are indeed looking at one structure, then the tethers and perhaps the ribs, likely belong to the newly emerging examples of thin magnetized threads linking larger regions of relativistic plasma -- i.e., an exciting new physical phenomenon.

\vspace{6pt}

\acknowledgments{We thank T.W. Jones, C. Nolting, C. Pfrommer , B. McKinley and Gopal-Krishna for interesting ideas on the origins of the features observed here and H. Akamatsu for providing the XMM image.}

\funding{LR's effort was funded in part by U.S. National Science Foundation grant AST17-14205 to the University of Minnesota.  MGCLS data products were provided by the South African Radio Astronomy Observatory and the MGCLS team and were derived from observations with the MeerKAT radio telescope. The MeerKAT telescope is operated by the South African Radio Astronomy Observatory, which is a facility of the National Research Foundation, an agency of the Department of Science and Innovation. K.Kn. is supported by the New Scientific Frontiers grant of the South African Radio Astronomy Observatory. }

\conflictsofinterest{The authors declare no conflict of interest.}


\reftitle{References}



\end{paracol}


%



\end{document}